\documentclass[fleqn,usenatbib]{mnras}

\usepackage{newtxtext,newtxmath}

\usepackage[T1]{fontenc}

\DeclareRobustCommand{\VAN}[3]{#2}
\let\VANthebibliography\thebibliography
\def\thebibliography{\DeclareRobustCommand{\VAN}[3]{##3}\VANthebibliography}

\usepackage{graphicx} 
\usepackage{amsmath} 
\usepackage{multirow} 
\usepackage{bookmark} 

\usepackage{caption}
\usepackage{threeparttable}

\title[Atmospheric Turbulence Profiling at Lenghu]{Atmospheric Turbulence Profiling in the Lenghu Site Area Using MERRA-2 Data}

\author[C. Yang et al.]{
Chen Yang,$^{1,2}$\thanks{E-mail: yc@stu.cwnu.edu.cn (CY)}
Licai Deng,$^{2,1}$\thanks{E-mail: licai@nao.cas.cn (LD)}
Kun Wang,$^{1}$\thanks{E-mail: kwang@cwnu.edu.cn (KW)}
Fei He,$^{3}$
Fan Yang,$^{2}$
Chunguang Zhang,$^{2}$
Xiaodian Chen,$^{2,1}$
\newauthor
Ruiyue Li,$^{2}$
and Linqing Tian$^{1}$
\\
$^{1}$School of Physics and Astronomy, China West Normal University, Nanchong 637009, China\\
$^{2}$National Astronomical Observatories, Chinese Academy of Sciences, Beijing 100101, China\\
$^{3}$Institute of Geology and Geophysics, Chinese Academy of Sciences, Beijing 100029, China
}

\pubyear{2026}

\begin{document}
\label{firstpage}
\pagerange{\pageref{firstpage}--\pageref{lastpage}}
\maketitle

\begin{abstract}

The Lenghu site, in the Saishiteng Mountains of the northern Tibetan Plateau, has been identified as an excellent astronomical site from preliminary statistics of clear nights, sky background brightness, precipitable water vapour, total seeing and weather. Since 2022, several projects with apertures larger than 4~m have started construction or are being planned. Given the use of adaptive optics (AO) on these projects, in-depth measurements and analysis of the atmospheric turbulence above the site are required: the distribution and variation of the turbulence profiles provide a large-scale, climatological basis for site planning and for comparing candidate locations, complementing rather than replacing the on-site measurements needed for instrument design and operations. Predicting the atmospheric profiles of the site and its surroundings from satellite data, combined with on-site measurements, can help to optimise the site planning. This paper uses MERRA-2 reanalysis data to study the atmospheric profiles of the site and its environment, revealing the seasonal and long-term variations of the observing conditions. The reanalysis results are cross-checked against on-site ground-level instruments serving the same purpose. MERRA-2 overestimates the median free-atmosphere seeing by $+0.22''$ ($+50\%$) while keeping the night-to-night correlation weak ($r\approx0.2$), and underestimates the median total seeing by $-0.24''$ ($-27\%$). It captures only $\sim$22\% of the ground-layer turbulence and little of the observed seasonal amplitude of the free atmosphere. MERRA-2 is therefore suited to relative comparisons of spatial variability between locations and to climatological characterisation, not to quantitative AO applications or night-by-night assessment of profiles.

\end{abstract}

\begin{keywords}
atmospheric effects -- turbulence -- site testing
\end{keywords}

\section{Introduction}

With its great potential in tackling atmospheric distortions for ground-based astronomical observations, the technique of adaptive optics (AO) has become a common practice for most new facilities, especially large-aperture telescopes and cutting-edge multi-conjugate instruments~\citep{hippler2019adaptive}. The site qualification requirements for the next generation optical telescopes have been reviewed in great detail by~\citet{schock2009tmt} in the case of the Thirty Meter Telescope (TMT), and~\citet{munoz2004siteselection} in the case of the Extremely Large Telescope (ELT). For the Lenghu site, many new projects, including several telescopes with apertures larger than 10~m, have been planned that demand more in-depth analysis of the atmospheric turbulence over the whole range of the site in order to better implement the AO technique, as commonly done for other major sites~\citep{hippler2019adaptive, masciadri2023optical}. The study of the atmospheric turbulence is very important to achieve the extreme scientific goals of these mega projects~\citep{hippler2019adaptive, masciadri2017optical, beltramo2020pushing}. The presence of atmospheric turbulence causes wavefront distortions, leading to star image jittering and blurring and therefore hinders the realization of design goals and investments of these facilities~\citep{fried1966optical}, while the measurements and analysis of the turbulence profile, i.e.\ the vertical distribution and seasonal variation of turbulence can help~\citep{joo2025one}. It is therefore crucial not only for evaluating the observing conditions of a site but also for planning the construction of large-aperture telescopes and science instruments~\citep{zhou2026atmospheric, feng2020site}. The Lenghu site has been tested preliminarily with conclusions on site qualities only for seeing-limited conditions~\citep{deng2021lenghu}, and there is an urgent need for a systematic study of the fine structure of atmospheric turbulence in the area above the site and its surroundings~\citep{chen2024astronomical}. This will be highly significant for the large-aperture telescope projects that are currently being fabricated and considered by many different institutions in China and international partners.

As a general practice for site selection, long-term monitoring and statistical studies of astronomical meteorological parameters are required at all professional observing sites such as Cerro Paranal in Chile~\citep{griffiths2024comparison, cuevas2024combining} and Mauna Kea in Hawaii~\citep{tokovinin2005optical, els2009thirty, chun2014extremely, vankooten2022climate}, forming a mature turbulence profile analysis system. By continuously monitoring these sites for decades, apart from integrated seeing (normally measured by a Differential Image Motion Monitor, DIMM), site monitoring has clarified the vertical layering characteristics of turbulence, such as the intensity of turbulence in the boundary layer and the distribution of turbulence in high-altitude jet stream regions, as well as established quantitative models for the seasonal and inter-annual variations of atmospheric parameter profiles. These results provide efficient references for the construction of large-aperture telescopes and the dynamic adjustment of observational strategies~\citep{masciadri2017optical, cuevas2024combining, li2026temperature}.

However, the preliminary research on similar parameters of the site is focused only on seeing-limited conditions, and is therefore still insufficient. Previous studies were mainly on the total turbulence intensities or relied on a few balloon experiments, lacking systematic analysis of the full vertical profile and long-term series. The long-term patterns of the primary astronomical meteorological parameters have been studied~\citep{zhao2022long, li2024cloud}, but not the vertical properties. Given that the Lenghu site will host modern large-aperture telescopes, the studies of these conditions require immediate actions.

Prior to this, our team has conducted studies on the integrated atmospheric turbulence, by measuring ground seeing using DIMM, at the Lenghu site~\citep{deng2021lenghu, chen2024astronomical}, and have analysed meteorological conditions using long-term multi-station observations~\citep{li2025characterization}. There is a lack of comparative analysis of turbulence in multiple large-scale spatial directions, and the analysis is limited to a single location during the first run of site testing. We have not systematically compared the influence of surrounding terrain in different directions on the vertical structure of turbulence, thus not providing refined references for selecting telescope placement.

The current study selects the Modern-Era Retrospective analysis for Research and Applications, Version 2 (MERRA-2) global reanalysis data, with a time span covering over 26 years from 2000 to 2026. This allows the large-scale seasonal evolution of the turbulence field to be followed over a long time series, complementing the monitoring programmes of established sites. The fields are provided on 72 model levels and a regular latitude--longitude grid, which resolves the large-scale structure of the turbulence field. The variables used are temperature and the horizontal wind components, which provide the input parameters for calculating $C_n^2$ and the derived coherence parameters~\citep{tatarski1961wave}; because the reanalysis resolves only the large-scale state, these quantities serve for relative comparisons between seasons and locations rather than as a substitute for on-site measurement of the ground-layer turbulence. In addition, we have also monitored the layered turbulent profile using Multi-Aperture Scintillation Sensor (MASS) developed by the Anhui Institute of Optics and Fine Mechanics, Chinese Academy of Sciences, starting from 2024 at the site, which provides an independent assessment to the major parameter profiles.

\section{The Area around the site} \label{sec:style}

\begin{figure*}
\centering
\includegraphics[width=\textwidth]{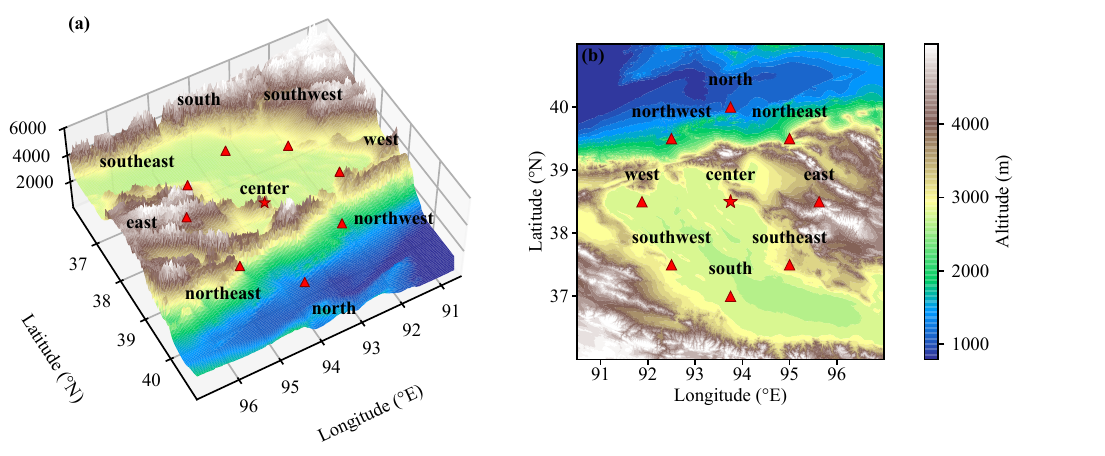}
\caption{Three-dimensional topographic map (a) and overhead projection view (b) of the wide region surrounding the Lenghu site, showing the nine selected locations.}
\label{fig1}
\end{figure*}

The Lenghu region is situated in the Qaidam Basin, near the northern edge of the Tibetan Plateau, characterised by an extremely arid climate with a typical highland desert landscape. It is cold and dry throughout the year, with slightly milder conditions in humidity and temperature during the summer season, and there is hardly any natural vegetation cover. Moreover, the region enjoys a favourable climate for astronomical observations, with clear skies both day and night. Figure~\ref{fig1} shows the three-dimensional topographic map (a) and overhead projection view (b) of the wide region surrounding the Lenghu site, in which Saishiteng Mountain rises sharply by about 1500 metres from the Qaidam Basin. The newly established observatory is located at the summit of the mountain, which serves as the central point of this study. Around this centre, we selected eight sampling locations distributed evenly in the directions of east, south, west, north, northeast, southeast, northwest and southwest, at distances of 155--167~km from the centre (mean 160 ~km), forming the grid for our analysis. The aim is to obtain the general trend and to compare the vertical aerodynamic properties over a wide range around the site.

\begin{figure*}
\centering
\includegraphics[width=0.8\textwidth]{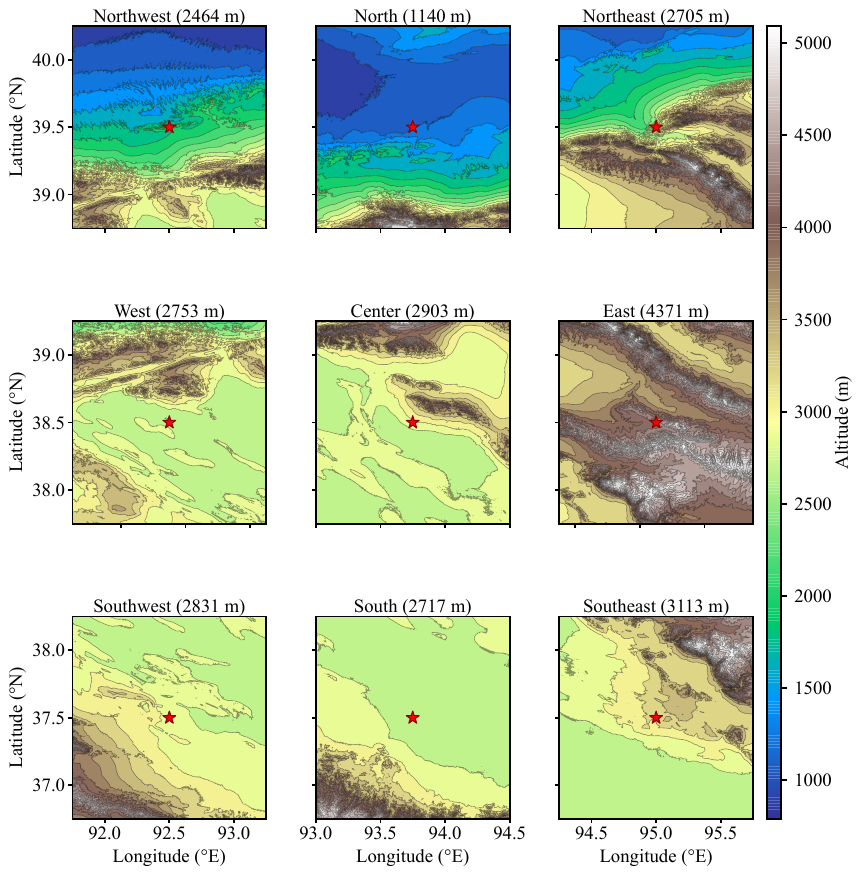}
\caption{Topographic maps of the vicinities of the nine selected observational sites. Each panel shows a $1.5^{\circ}\times1.5^{\circ}$ window centred on the MERRA-2 grid point, with the site name and the corresponding SRTM terrain elevation printed above it; the tick numbers are given on the outer frame only and the colour bar gives the elevation in metres. The geographic coordinates of the nine grid points are listed in Table~\ref{tab:grid_points}.}
\label{fig_2}
\end{figure*}

The detailed topographic maps of these nine locations are shown in Figure~\ref{fig_2}. The three northern points are located in the Suoerkuli Valley at the northern foot of the Altun Mountains, where the altitude is the lowest across the entire study area, characterised by a low-lying central depression flanked by gentle hills on either side. The three southern points — southwest, south and southeast — are situated on the pre-mountain plateau of the Qimantag Mountains. The central point and the eastern point both belong to the Saishiteng Mountains, where the terrain rises steeply from west to east; the central point sits on a gentle highland at the foot of the mountain, whilst the eastern point is the highest ridge peak in the entire study area. The geographic information of the nine selected points are listed in Table~\ref{tab:grid_points}, in which the coordinates give the grid-centre positions used in the analysis. The elevations are the real terrain height at each grid point taken from the Shuttle Radar Topography Mission (SRTM) digital elevation model, and the surface pressure is derived from the elevation using the standard atmosphere. These values document the terrain of the smoothed model grid; they are not used as the lower limit of the turbulence integration, which is taken at the common Platform C site level of 4.2~km (600~hPa) for all nine sites, so that the same atmospheric layer is compared throughout (for the MASS-based quantities of Table~\ref{keystats} the limit is instead the 562.5~hPa lower boundary of the MASS measurements). Note that the centre grid point (2902~m) lies below the actual Platform C site (4200~m), because the $\sim$55~km MERRA-2 grid smooths over the mountain peak.

\begin{table}
\centering
\caption{Locations, real terrain elevations and corresponding surface pressures of the nine selected grid points (near Lenghu). \label{tab:grid_points}}

\begin{tabular}{rllrr}
\hline
\# & Direction & Geo. Coords. & Elevation & Surface pressure\\
&&&(m)&(hPa)\\
\hline
1 & centre & $38.5^{\circ}$N, $93.75^{\circ}$E & 2902 & 710 \\
2 & north & $40.0^{\circ}$N, $93.75^{\circ}$E & 1140 & 884 \\
3 & northeast & $39.5^{\circ}$N, $95.0^{\circ}$E & 2705 & 728 \\
4 & east & $38.5^{\circ}$N, $95.625^{\circ}$E & 4345 & 589 \\
5 & southeast & $37.5^{\circ}$N, $95.0^{\circ}$E & 3113 & 691 \\
6 & south & $37.0^{\circ}$N, $93.75^{\circ}$E & 2716 & 727 \\
7 & southwest & $37.5^{\circ}$N, $92.5^{\circ}$E & 2834 & 716 \\
8 & west & $38.5^{\circ}$N, $91.875^{\circ}$E & 2753 & 723 \\
9 & northwest & $39.5^{\circ}$N, $92.5^{\circ}$E & 2453 & 751 \\
\hline
\end{tabular}
\end{table}

\section{Data Sources}
\label{sec:style_2}

The meteorological fields used in this study come from the National Aeronautics and Space Administration (NASA) Global Modeling and Assimilation Office (GMAO) MERRA-2 product~\citep{gelaro2017merra2}. MERRA-2 uses the Goddard Earth Observing System (GEOS) model, version 5.12.4,~\citep{molod2015development} to initialise the forecast fields and the Global Statistical Interpolation (GSI) scheme to fuse them optimally with satellite observations including Terra/Aqua, the Geostationary Operational Environmental Satellite (GOES) series and the Constellation Observing System for Meteorology, Ionosphere and Climate (COSMIC), as well as ground-based and conventional observations after quality control, bias correction and interpolation; the Incremental Analysis Update (IAU) technique then ensures a smooth transition to the next forecast integration. The resulting globally homogenised product has been generated continuously since 1980 and is distributed through NASA's Goddard Earth Sciences Data and Information Services Center (GES DISC; https://disc.gsfc.nasa.gov/).

This work uses the instantaneous three-dimensional assimilated meteorological field product \texttt{inst3\_3d\_asm\_Nv} (``inst'' = instantaneous, ``3d'' = three-dimensional, ``asm'' = assimilated, ``Nv'' = product identifier). It has a 3-hour temporal resolution, i.e.\ eight steps per day from 00:00~UTC (Coordinated Universal Time), and a 72-layer vertical grid~\citep{bosilovich2015merra2}, with a horizontal resolution of $0.5^\circ$ (latitude) $\times$ $0.625^\circ$ (longitude). At the latitude of the Lenghu site ($\sim 38.5^\circ$N), this grid spacing corresponds to approximately $55.6\times54.4$~km$^2$, which is far coarser than the thickness of the ground turbulence layer (typically a few hundred metres) and therefore cannot resolve the local topography of the mountain peak; this is a key limitation for characterising the ground layer, as discussed in Section~\ref{comparisons}. The product provides temperature, pressure, humidity and wind fields in Network Common Data Form version 4 (NetCDF-4) format; the variables used here are the atmospheric temperature $t$ and the two horizontal wind components $u$ and $v$.

The time span selected for this study is from January 2000 to May 2026. The 6th time index, corresponding to 18:00 UTC (2:00 AM Beijing Time), is chosen as it is close to midnight local time; unless stated otherwise, this snapshot is used as the nighttime representative for all results presented below, and any analysis that uses a different time is identified explicitly where it occurs. The selected longitude and latitude ranges are from $90.0^\circ$E to $95.625^\circ$E, and from $35.0^\circ$N to $42.0^\circ$N. The variables used in this study include atmospheric temperature $(t)$ and the two vertical components of horizontal wind speed $(u, v)$. The data dimensions and variables are summarized in Table~\ref{tab:merra2_dims}. For the nine-site analysis in the Lenghu region this 18:00 UTC snapshot is used throughout, whereas for the inter-site comparison of Table~\ref{keystats} the snapshot closest to local midnight is adopted at each site, as described in Section~\ref{comparisons}.

\begin{table}
\centering
\caption{MERRA-2 product dimensions and variables used in this study. \label{tab:merra2_dims}}
\begin{tabular}{rlll}
\hline
Number & Units & Label & Explanation\\
\hline
1 & --- & time & minutes since 1999-01-01 \\
2 & degrees\_east & lon & longitude \\
3 & degrees\_north & lat & latitude \\
4 & layer & lev & model vertical level \\
5 & K & t & air temperature \\
6 & m/s & u & eastward wind \\
7 & m/s & v & northward wind \\
\hline
\end{tabular}
\end{table}

\section{Methods and data processing}
\label{sec:style_3}

 Before the individual quantities are defined, the limitations that follow from the nature of the reanalysis data should be stated, because they determine how the results of this study can be used. The MERRA-2 fields are provided on a horizontal grid of $\sim$55~km and on 72 model levels sampled every three hours: the vertical spacing is therefore coarser than the thickness of the thin turbulent layers that dominate the optical effects, and the horizontal resolution cannot represent the topography of an isolated mountain peak. Consequently, the profiles derived here describe the large-scale, climatological state of the atmosphere above the region and are suited to relative comparisons between seasons and between locations, whereas the ground-layer turbulence, which is generated by the unresolved topography, cannot be recovered from them and must be measured directly on site, as quantified in Section~\ref{comparisons}.

 \subsection{Calculation of the Atmospheric Refractive Index Structure Constant} \label{subset:tables}

In the atmosphere, turbulence causes mixing and heat exchange of all contents, including thermal and dynamical energies, between air masses, leading to small fluctuations in local temperatures. The amplitude and frequency of these temperature fluctuations are closely related to the intensity of atmospheric turbulence. In actual calculations, the two-point temperature difference structure function method is used, employing micro-temperature sensings to detect local temperature changes $\Delta T$. The environmental temperature change is converted into a change in resistance, which is then converted into a voltage change $\Delta U$ through an unbalanced bridge and finally amplified by a voltage amplifier.
\begin{equation}
\Delta U = A \cdot \Delta T
\end{equation}

In the known temperature field, the temperature structure function between two points is defined as the statistical average of the square of the temperature difference between the two points:
\begin{equation}
D_T(\vec{r}) = \langle [T(\vec{r}_1) - T(\vec{r})]^2 \rangle
\end{equation}
According to the local isotropic assumption in Kolmogorov's turbulence theory, the temperature structure constant can be obtained:
\begin{equation}
C_T^2 = D_T(\vec{r}) \cdot r^{-2/3}
\end{equation}

A single-point temperature structure function method can also be used, with a similar principle. At a selected single point, high-precision temperature sensors are used to collect temperature data at regular intervals (e.g., once per second). The collection time usually lasts for several hours or even longer to ensure statistical validity. The collected temperature data are then processed according to the definition of the temperature structure function.
\begin{equation}
C_T^2(h) = \langle [T(t) - T(t - \tau)]^2 \rangle (V_\tau)^{-2/3},
\end{equation}

where, $\Delta t$ represents the time interval, and the spatial length is typically set to 1 metre. The magnitude of $\Delta t$ is determined by the measured mean wind speed.
The refractive index field exhibits properties similar to those of the temperature field. Likewise, the relationship between the refractive index structure function and the refractive index structure constant can be derived accordingly.
\begin{equation}
D_n(\vec{r}) = \langle [n(\vec{r}_1) - n(\vec{r})]^2 \rangle = C_n^2(\vec{r})^{2/3}
\end{equation}

Fluctuations in temperature lead to corresponding fluctuations in the refractive index. Under locally homogeneous and isotropic turbulence conditions, and neglecting the influence of humidity, the correlation between the refractive index and temperature $T$ as well as pressure $P$ are considered. For the visible and near-infrared spectral bands, according to the Tatarskii formula ~ \citep{tatarski1961wave}, the relationship between the refractive index structure constant and the temperature structure constant can be expressed as follows:
\begin{equation}
C_n^2(h) = \left(7.9 \times 10^{-5} \frac{P}{T(h) \cdot \theta(h)}\right)^2 \cdot C_T^2(h),
\end{equation}

where the unit of pressure is hectopascal (hPa), and the unit of temperature is Kelvin (K). In addition to being obtained through measurements using microthermal sensors, the temperature structure constant can also be expressed in an alternative form,
\begin{equation}
C_T^2(h) = c_0 L(h)^{4/3} \left[ \frac{\partial \theta(h)}{\partial h} \right]^2,
\end{equation}

where $c_0$ is a constant, and $L(0)$ represents the outer scale, satisfying the following relationship:
\begin{equation}
L(h) = \sqrt{\frac{2S(h)^2}{\frac{g}{\theta(h)} \cdot \frac{\partial \theta(h)}{\partial h}}},
\end{equation}

where $S$ denotes the wind shear and $\theta$ represents the potential temperature. They satisfy the following relationships:
\begin{equation}
S = \sqrt{\left( \frac{\partial u}{\partial h} \right)^2 + \left( \frac{\partial v}{\partial h} \right)^2},
\end{equation}

and

\begin{equation}
\theta(h) = T(h) \cdot \left[ \frac{1000}{P(h)} \right]^{0.286}
\end{equation}

Here, $u$ and $v$ are the two orthogonal components of the horizontal wind velocity.

Finally,

\begin{equation}
C_n^2(h) = 8.23 \times S(h)^{4/3} \cdot \theta(h)^{2/3} \left[ \frac{7.9 \times 10^{-5} P(h)}{T(h) \cdot \theta(h)} \right]^2 \cdot \left[ \frac{\partial \theta(h)}{\partial h} \right]^{4/3}.
\end{equation}

Consequently, the refractive index structure constant can be derived from standard meteorological parameters.

 The parameterisation is proportional to the $4/3$ power of the wind shear and of the potential-temperature gradient, $C_n^2 \propto S^{4/3}(\partial\theta/\partial h)^{4/3}$; it is non-negative and vanishes as either gradient vanishes. In the very weakly turbulent layers of the upper stratosphere and mesosphere the resolved gradients are themselves small, so the formula returns correspondingly tiny values, and those gradients are in addition dominated by noise, so the values cannot be regarded as a valid turbulence measurement. In all the column integrals, which run from the site level (600~hPa) to the top of the reanalysis column ($0.01$~hPa, $\approx76$~km), we therefore retain only those layers with $10^{-17.5} \le C_n^2 \le 10^{-14.5}~\mathrm{m}^{-2/3}$, which is equivalent to treating the layers outside that range as laminar. The discarded layers lie in the upper stratosphere and mesosphere (median height $\approx40$~km) and carry negligible weight in the column integral of $C_n^2$, but not in the height-weighted integral that defines the isoplanatic angle (Section~\ref{subsec:theta0}). The selection is therefore not entirely without consequence: for the daily profiles of 2000--2026 it lowers the free-atmosphere seeing and raises $r_0$ by $1.2\%$ each, and raises $\theta_0$ by $6.1\%$, all in the median and all relative to retaining every layer, and the values reported below refer to this convention. The upper limit of the retained range is never reached above the site level and has no effect.

 \subsection{The Atmospheric Coherence Length} \label{subsec:tables_2}

In ground-based optical astronomical observations, multiple parameters are often used together to evaluate the atmospheric stability at a site. These parameters, such as coherence length, isoplanatic angle, coherence time, full-width at half-maximum (FWHM), and Richardson number, are all derived based on the atmospheric refractive index structure constant through rigorous mathematical integration and derivation.
Coherence length, also known as the Fried constant ~ \citep{fried1966optical}, refers to the distance within which a wavefront remains relatively smooth as the light wave propagates through the atmosphere. In common interference phenomena, within this distance, the light wave always has the ability to produce clear interference fringes, allowing the interference phenomenon to be observed normally. This distance is what the coherence length represents. The coherence length $r_0$ is inversely related to seeing: a smaller $r_0$ corresponds to worse seeing. According to Fried's theory, the coherence length is related to the atmospheric refractive index structure constant as follows:

\begin{equation}
r_0 = \left[ 0.423 \kappa^2
 J \right]^{-3/5}, \qquad J \equiv \int_0^L C_n^2(h) dh \label{eq:r0}
\end{equation}

where $\kappa$ is the wave number, $\lambda$ is the wavelength, $L$ is the propagation path length, and $J$ is the turbulence integral. The larger $J$, the shorter the coherence length, meaning that atmospheric turbulence causes a greater disruption to the coherence of the light wave.

 \subsection{The Isoplanatic Angle}
\label{subsec:theta0}

When adaptive optics is used to correct the wavefront, the correction is derived from a guide star and becomes increasingly inaccurate for targets separated from it. The isoplanatic angle is the angular separation over which two wavefronts share essentially the same turbulence induced phase distortion, that is, the field of view over which a single correction remains valid. It is obtained from the same integral of the atmospheric refractive index structure constant as the coherence length, but with each layer weighted by its height above the aperture:
\begin{equation}
\theta_0 = \left[ 2.91\kappa^2\int C_n^2(h)\,h^{5/3}\,dh \right]^{-3/5}
\end{equation}

where $\kappa$ is the wave number and $h$ is the height above the aperture. The $h^{5/3}$ weighting makes $\theta_0$ considerably more sensitive to the vertical distribution of the turbulence than the seeing is: layers in the upper troposphere and the lower stratosphere contribute little to $\int C_n^2(h)dh$ but dominate the integral above, so that a thin high layer can leave the seeing almost unchanged while substantially reducing the isoplanatic angle.

 \subsection{The Coherence Time}
\label{subsec:tables_3}

Coherence time is used to measure how long a light wave can maintain its coherence. It is derived from the Taylor frozen turbulence hypothesis and related theoretical derivations:
\begin{equation}
\tau_0 = 0.314\left[0.423\kappa^2\int C_n^2(h)\cdot V^{5/3} dh\right]^{-3/5} = 0.314\frac{r_0}{\overline{V}}
\end{equation}

\subsection{ The Full Width at Half Maximum (FWHM)}
\label{subsec:tables_4}

When light passes through the atmosphere and forms an image at the telescope focal plane, the resulting image is often a blurred spot. The FWHM is used to quantify the degree of this blurring, specifically the width of the intensity distribution at half of the peak value. This is the classic definition of seeing. From the point spread function perspective, FWHM is closely related to the atmospheric refractive index structure constant. For imaging under the Kolmogorov turbulence model, FWHM is complexly related to coherence length and other parameters, essentially affected by the integral of the atmospheric refractive index structure constant. The larger the atmospheric refractive index structure constant, the more the image spot widens, meaning the FWHM becomes larger, and the imaging resolution significantly decreases.

\begin{equation}
\varepsilon_{FWHM} = 5.25\lambda^{-1/5}J^{3/5} = 0.98\frac{\lambda}{r_0}
\label{eq:seeing}
\end{equation}

For the same wavelength of light, FWHM is inversely proportional to the coherence length. The stronger the integral effect, the shorter the coherence length, and the wider the FWHM, indicating that the image of the target object becomes more diffuse. Since both $r_0$ and $\varepsilon_{FWHM}$ depend on the turbulence integral only through $J$ (Equations~\ref{eq:r0} and~\ref{eq:seeing}), at a fixed wavelength $\varepsilon^{5/3}$ is proportional to $J$; because $J$ is additive along the line of sight, the seeing produced by two independent layers adds in the $5/3$ power.
The atmospheric refractive index structure constant plays a key role in assessing the atmospheric stability at the site. Through a series of rigorous and varied calculation methods, it is possible to quantify the degree to which atmospheric optical turbulence affects ground-based optical astronomical observations. Many important parameters, such as coherence length, isoplanatic angle, coherence time, FWHM, and Richardson number, all heavily depend on the calculation and derivation of the atmospheric refractive index structure constant. These derived parameters characterise the optical properties, turbulence characteristics, and stability of the atmosphere from different dimensions.

\section{Results}
\label{sec:style_4}

The next step is to provide a detailed characterisation of the profiles of various parameters and analyse the vertical distribution characteristics and temporal evolution of atmospheric turbulence in the Lenghu region.

The grid point closest to Platform C, located at 93.75$^\circ$E and 38.5$^\circ$N, was selected as the centre point. Eight additional grid points were selected in different directions at distances of 155--167~ km from this centre point.

 \subsection{Temperature and Wind Field Profiles at Various Directions around Platform C}
\label{subsec:tables_5}

Atmospheric turbulence is a key factor determining the observing quality of an astronomical site, governed by both thermal and dynamic mechanisms. Vertical temperature gradients induce stratification instability and generate thermal turbulence, while vertical wind shear excites dynamic turbulence. The spatiotemporal coupling of these two processes jointly determines the strength and distribution of turbulence throughout the atmospheric column. Therefore, understanding the vertical and seasonal evolution of temperature and wind fields at a site provides the foundation for quantitative analysis of atmospheric turbulence characteristics. In the following, the temperature and wind field structures at the Lenghu Platform C and its surrounding directional points are analysed to reveal the modulating effects of local topography on thermal and dynamic conditions.

Height-layered temperature and wind speed data were selected for the Platform C grid point ($93.75^\circ$E, $38.5^\circ$N) and then time-averaged; the resulting temperature and wind field profiles at this point are plotted below.

Figure~\ref{fig:general_3} presents the monthly mean vertical temperature structure at Platform C. The lower and middle troposphere exhibit a pronounced seasonal cycle with a warm core in summer driven by strong surface heating and cold conditions in winter; in contrast, the upper troposphere and lower stratosphere (above 100~hPa) remain cold throughout the year. The World Meteorological Organization (WMO) thermodynamic tropopause follows this seasonality, rising to approximately 100~hPa in summer and descending to 180--200~hPa
in winter. The four-season profiles (panel b) show that the seasonal temperature contrast is largest near the surface and persists upward into the middle troposphere, while all seasons converge above the tropopause.

\begin{figure*}
\centering
\includegraphics[width=0.8\textwidth]{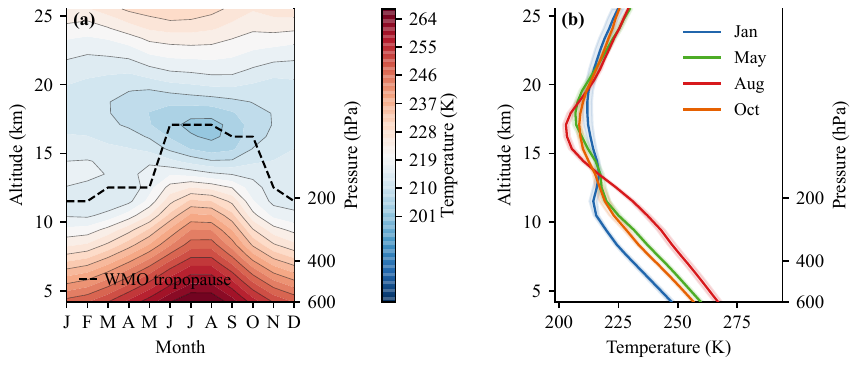}
\caption{Monthly mean heatmap of the nighttime temperature field from 2000 to 2026 (a) and four-season vertical profile (b); the dashed line in the heatmap of panel (a) marks the WMO thermal tropopause. The left vertical axis of both panels is the altitude above sea level on a uniform scale, and the right axis gives the corresponding pressure.}
\label{fig:general_3}
\end{figure*}

Overall, the vertical temperature structure at Platform C exhibits characteristics typical of a mid-latitude plateau site: the lower troposphere is strongly modulated by surface thermal forcing, with a prominent warm-summer/cold-winter contrast, whereas above 100~hPa in the lower stratosphere, local heating effects are weak and temperature shows little seasonal fluctuation. Driven by the seasonal circulation, the WMO thermodynamic tropopause height follows an annual cycle of summer ascent and winter descent. Taken together, the seasonal thermal contrast in the lower atmosphere is far stronger than in the stratosphere.

The generation and evolution of atmospheric turbulence are jointly controlled by thermal stratification and atmospheric dynamic wind fields; therefore, the vertical wind field characteristics at Platform C are further analysed.

\begin{figure*}
\centering
\includegraphics[width=0.8\textwidth]{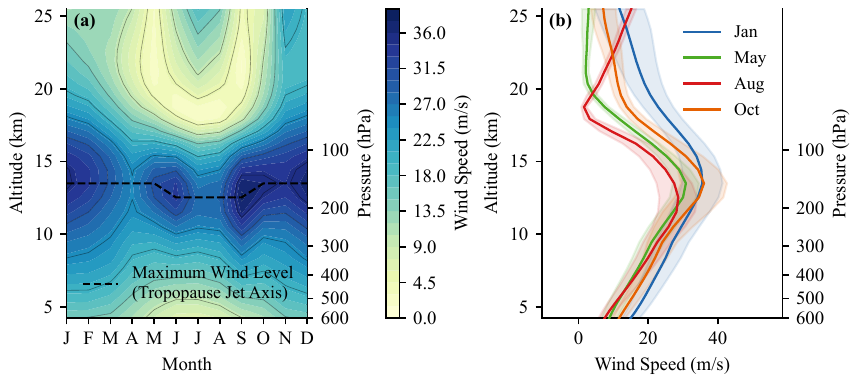}
\caption{Monthly mean heatmap of the nighttime wind speed field from 2000 to 2026 (a) and four-season vertical profile (b); the dashed line in the heatmap of panel (a) marks the level of maximum wind speed, i.e.\ the tropopause jet axis. Similar to Fig.~\ref{fig:general_3}, the left vertical axis is the altitude above sea level on a uniform scale, and the right vertical axis gives the corresponding pressure.}
\label{fig:general_5}
\end{figure*}

Figure~\ref{fig:general_5} presents the monthly mean wind structure at Platform C. Near-surface winds are weak throughout the year and increase with height to a maximum at the tropopause, where the mid-latitude westerly jet dominates. The jet is strongest and highest in winter (peak approximately 35~m/s near 150--200~hPa), and weakest and lowest in summer (approximately 25~m/s near 200~hPa), when the South Asian monsoon weakens and shifts it northward.

The wind and temperature fields at the Lenghu Platform C together form the complete atmospheric thermodynamic--dynamic background: near-surface winds are persistently weak throughout the year with small seasonal differences; the westerly jet in the middle-to-upper troposphere is strong in winter and weak in summer, and high in winter and low in summer in terms of altitude. The southward shift of the westerly belt in winter strengthens and elevates the jet, while the summer monsoon and plateau heating jointly weaken and depress it; the variation of the upper-level wind field is primarily driven by the tropopause jet. Temperature gradients generate thermal turbulence, while wind shear in the upper-level jet produces dynamic turbulence; the seasonal coupling of these two factors governs the distribution of atmospheric turbulence. These temperature and wind fields represent the synoptic-scale background resolved by the $\sim$55~km MERRA-2 grid, however; the local ground-layer turbulence generated by the unresolved topography is not captured here and is quantified separately in Section~\ref{comparisons}.

 \subsection{Atmospheric Refractive Index Structure Constant in Lenghu}
\label{subsec:tables_6}

After analysing the vertical distribution and seasonal characteristics of the temperature and wind fields in the Lenghu region, the atmospheric refractive index structure constant becomes the key parameter for describing the degree of atmospheric turbulence's effect on light propagation. The larger its value, the stronger the turbulence. As described above, the atmospheric refractive index structure constant is related to the atmospheric temperature gradient and wind speed shear. Combining the previous evolution patterns of the temperature field and wind speed field, the vertical distribution and seasonal variation characteristics of the atmospheric refractive index structure constant will be further explored.

\subsubsection{Seasonal Characteristics of the Vertical Structure of Atmospheric Turbulence at Platform C}
\label{subsec:tables_6_1}

At present, Platform C at Lenghu is a relatively well-developed observing platform, with several telescopes already in operation. Before conducting a multi-point spatial comparison across the entire region, it is necessary to first clarify the background characteristics of atmospheric turbulence at the central Platform C itself.

\begin{figure*}
\centering
 \includegraphics[width=\textwidth]{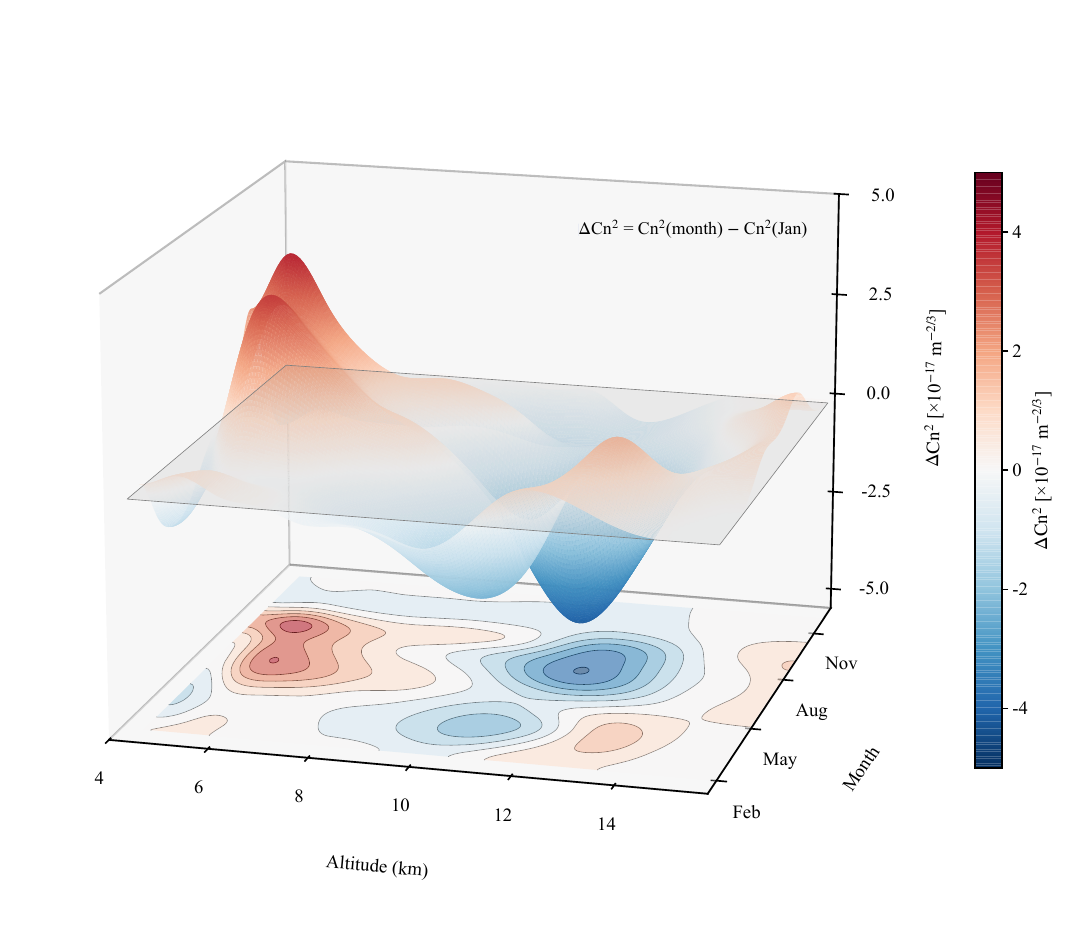}
 \caption{Three-dimensional surface plot of the seasonal $C_n^2$ difference for nighttime at Platform C from 2000 to 2026. }
\label{fig:general_9}
\end{figure*}

Figure~\ref{fig:general_9} uses the same nighttime dataset as Fig.~\ref{fig:general_3}: the temperature and wind profiles were used to calculate the atmospheric refractive index structure constant $C_n^2$. The $C_n^2$ values at each pressure level were then averaged by month. January was selected as representative of winter, and the $C_n^2$ values of the remaining months were subtracted from those of January to analyse the seasonal variation of the vertical structure of atmospheric turbulence at Platform C. Only the layers above the site level (4200~m) are retained, so that the profiles refer to the same air column that the site instrumentation samples.

In Figure~\ref{fig:general_9}, the horizontal axis represents the altitude, and the vertical axis takes January as the winter reference and subtracts it from the remaining months. The height of the three-dimensional surface represents the turbulence difference between each month and winter, with positive values (warm colours) indicating stronger turbulence activity than in winter, and negative values (cool colours) indicating weaker turbulence than in winter.

 In the lowest layer the deviations are small and predominantly positive; in the middle troposphere at 400--500~hPa and at the tropopause near 200~hPa, in contrast, completely opposite seasonal response characteristics are observed.

The atmospheric turbulence at Platform C exhibits significant warm--cold season differences. For conventional optical observations, where imaging quality is primarily affected by turbulence in the middle troposphere at 400--500 hPa, the nighttime turbulence intensity in spring and summer is significantly higher than in winter, resulting in poorer imaging performance. In contrast, for high-resolution observing systems such as adaptive optics, whose correction effectiveness is highly dependent on the atmospheric state at the 200 hPa tropopause level, the turbulence disturbances at this altitude in summer and autumn are far weaker than in winter, making these seasons more favourable for high-resolution astronomical observations.

\begin{figure}
\centering
 \includegraphics[width=0.9\columnwidth]{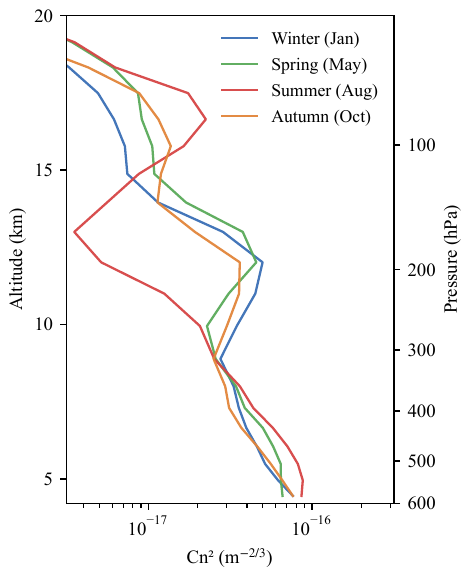}
 \caption{Four-season vertical profile of nighttime $C_n^2$ at Platform C from 2000 to 2026. }
\label{fig:general_10}
\end{figure}

 To more intuitively compare the vertical distribution differences of nighttime atmospheric turbulence across the four seasons at the central Platform C in Lenghu, the $C_n^2$ profiles of the four typical seasons are plotted together for comparison, as shown in Figure~\ref{fig:general_10}. In the figure, the vertical axis represents the altitude, the right axis gives the corresponding pressure, and the horizontal axis represents the refractive index structure constant $C_n^2$. Overall, the turbulence is stronger in the near-surface layer, and the four seasonal profiles show pronounced seasonal differences, most notably in summer, corroborating the earlier Figure~\ref{fig:general_9}.

\begin{figure*}
\centering
 \includegraphics[width=\textwidth]{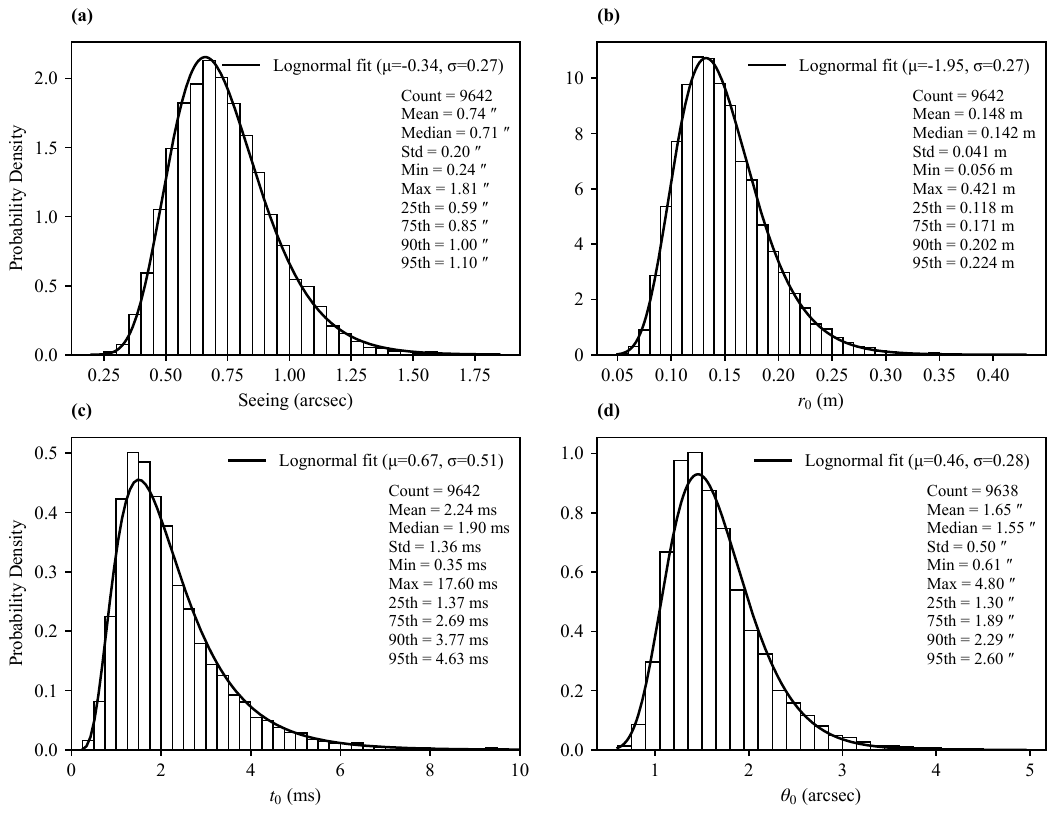}
 \caption{Statistics of atmospheric turbulence related parameters. (a) Free-atmosphere seeing; (b) coherence length $r_0$; (c) coherence time $t_0$, with the horizontal axis truncated at $10$~ms, beyond which only $0.3\%$ of the sample lies; and (d) isoplanatic angle $\theta_0$.}
\label{fig:keystats}
\end{figure*}

 The free-atmosphere seeing sets the overall image quality, the coherence length $r_0$ the actuator spacing and hence the order of the correction, the isoplanatic angle $\theta_0$ the corrected field of view, and the coherence time $t_0$ the bandwidth required of the control loop. Figure~\ref{fig:keystats} presents the statistics of these four quantities above the site, all integrated above the site level and evaluated at $\lambda=500$~nm.
 For telescopes working at Platform C, the free-atmosphere seeing derived here is systematically lower than the full-column DIMM statistics from the ongoing monitoring (http://lenghu.china-vo.org) and from previous publications~\citep{deng2021lenghu}, as expected because it excludes the ground layer; the ground-layer contribution itself is quantified in Section~\ref{comparisons}.

\subsubsection{Spatial Variability of Atmospheric Turbulence in the Lenghu Region}
\label{subsec:tables_6_2}

\begin{figure*}
\centering
 \includegraphics[width=\textwidth]{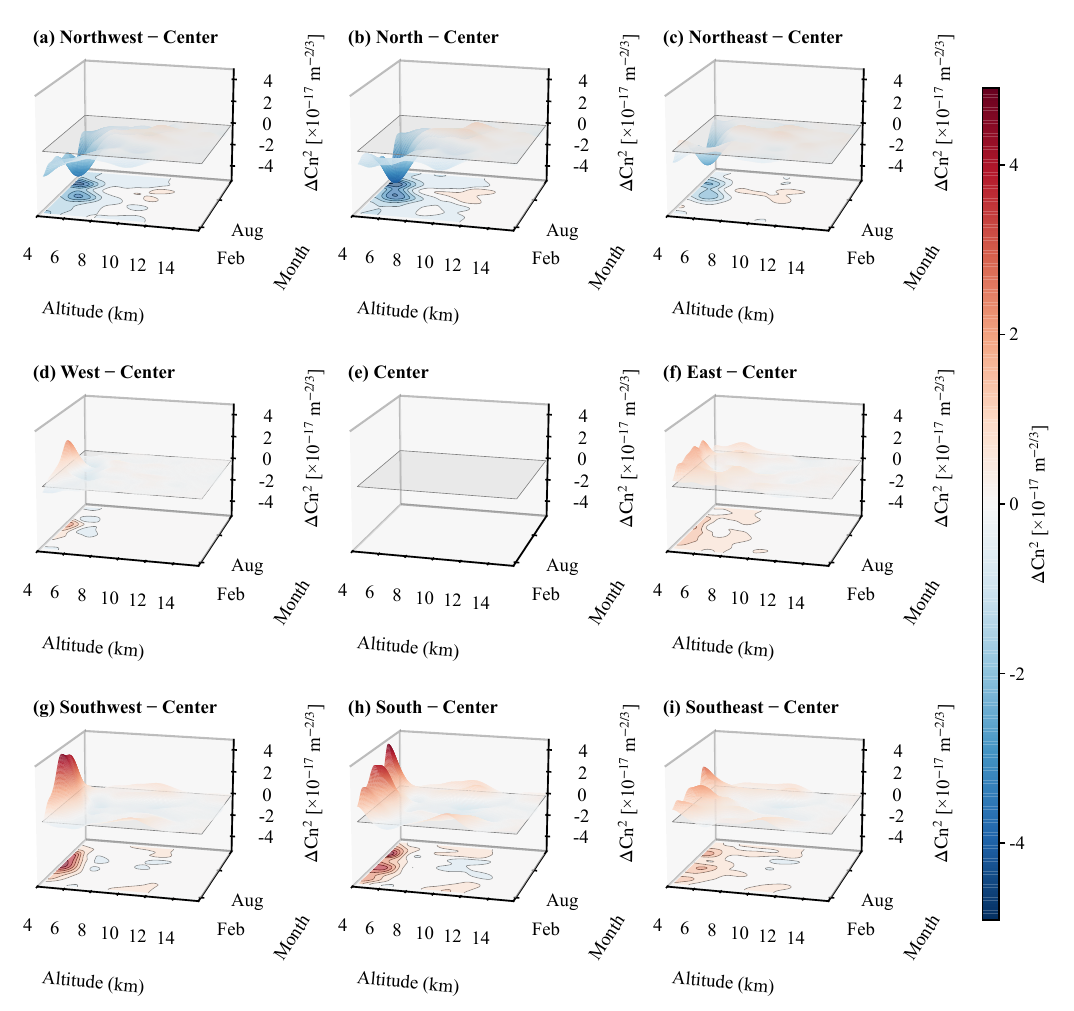}
 \caption{Three-dimensional surface plot of the nighttime $C_n^2$ difference between each directional site and the central point from 2000 to 2026. The month axis is labelled at February and August only, for legibility at this panel size.}
\label{fig:general_11}
\end{figure*}

However, different observation sites in the Lenghu region are influenced by local topography and near-surface thermal processes, leading to pronounced spatial differentiation in the turbulence environment. This section takes the central Platform C as the baseline and compares the differences in the vertical structure of atmospheric turbulence between the surrounding sites and the central point across different time periods, aiming to identify the spatial distribution differences of atmospheric turbulence in the region and further refine the complete picture of regional atmospheric turbulence. For the other 8 locations surrounding the central point in the Lenghu region, as described in table \ref{tab:grid_points} in section \ref{sec:style}, the deviations related to the atmospheric refractive index structure constant of Platform C are calculated. The results are presented in Figure~\ref{fig:general_11}.

 Similar to Figure~\ref{fig:general_9} presented earlier, the horizontal axis represents the altitude above sea level, and the vertical axis covers the full year from January to December.

The profile characteristics across all eight directions exhibit a clear north--south differentiation pattern and can be classified into two characteristic regions: the three northern sites (northwest, north, northeast) and the five southern and eastern sites (west, east, southwest, south, southeast). These two groups display completely opposite turbulence deviation patterns. The three northern subplots are dominated by large areas of dark blue negative values, with the negative deviation core concentrated in the middle and lower troposphere at 400--600 hPa. For the vast majority of months throughout the year, $\Delta C_n^2$ in this layer is negative; only at the 200 hPa upper level and during a few autumn and winter months do weak warm-colour positive deviations appear, indicating that upper-level turbulence is slightly higher than at the central site only over limited periods and altitude ranges. This indicates that the turbulence in the middle and lower layers on the northern side is overall weaker than at the central Platform C, with the negative deviation amplitude reaching its annual peak during the spring and summer convective period, making the advantage for conventional optical observations most pronounced. In sharp contrast, the subplots for the east, south, southwest, and southeast directions are dominated by large-scale warm-colour positive deviations, similarly concentrated in the lower troposphere at 500--600 hPa. Positive values prevail throughout the year, with red bulges reaching their maximum during the spring and summer months from May to August, indicating that the near-surface and mid-level atmospheric turbulence in the southeast and south sectors is consistently higher than at the central platform. Among these, the surface for the due-west site remains close to the zero baseline overall, with only a weak positive deviation in the lower layers during spring and summer, making it the transitional orientation within the southern sector where the turbulence environment differs least from the central site.

From the perspective of common vertical stratification patterns, the middle and lower troposphere at 600--400 hPa is the layer where the spatial differences among all sites are most pronounced, with all of the north--south contrast in observing conditions concentrated in this layer. In contrast, at the 200 hPa tropopause level, the turbulence differences across all eight directions approach zero throughout the year, indicating that within this 155--167~ km spatial range, the large-scale upper-level circulation is highly uniform, and tropopause-level turbulence exhibits almost no spatial variability. For high-resolution adaptive optics observations that are sensitive to the tropopause, there is no significant site-to-site distinction across the entire study area.

Examining the seasonal evolution, during the late winter and late autumn months (January--February and November--December), the difference surfaces in all subplots remain close to the zero baseline. At these times, the turbulence differences between each site and the central platform are significantly reduced, and the regional spatial heterogeneity is at its weakest. In contrast, during the spring and summer months from May to August, the warm- and cool-colour deviation extremes are simultaneously amplified, and the north--south contrast in turbulence intensity reaches its annual maximum, making this the period when the spatial differentiation of observing conditions across the entire Lenghu region is most pronounced.

In summary, the atmospheric optical turbulence in the Lenghu region exhibits a pronounced north--south asymmetric spatial distribution: the observing conditions in the middle and lower layers at all northern sites are generally better than those at the central Platform C, while the near-surface and mid-level turbulence disturbances at the southern and eastern sites are consistently stronger throughout the year. Local topography and near-surface thermal circulation dominate the spatial differences in the middle and lower troposphere, whereas the upper-level tropopause is controlled by large-scale circulation, resulting in regionally uniform turbulence.

\subsubsection{Seeing in the Lenghu Region}
\label{subsec:tables_6_3}

The previous sections have compared the spatial and seasonal differences of layer-resolved turbulence $C_n^2$ between the central Platform C and the eight surrounding sites, which reflect the turbulence intensity at individual pressure levels. In practical observations, seeing is commonly used to directly characterise the actual imaging performance of a telescope, as seeing represents the integrated result of atmospheric turbulence over the entire atmospheric column.

By integrating the nighttime $C_n^2$ vertical profiles over multiple years at the nine sites, the monthly mean seeing curves can be calculated. It should be noted that this seeing is computed by integrating atmospheric turbulence only above the site level, so that the same atmospheric layer is compared across all nine sites. This metric therefore characterises only the free-atmosphere contribution to imaging degradation, and serves as a basis for horizontally comparing the high-altitude observing conditions among the different sites.

\begin{figure*}
\centering
 \includegraphics[width=\textwidth]{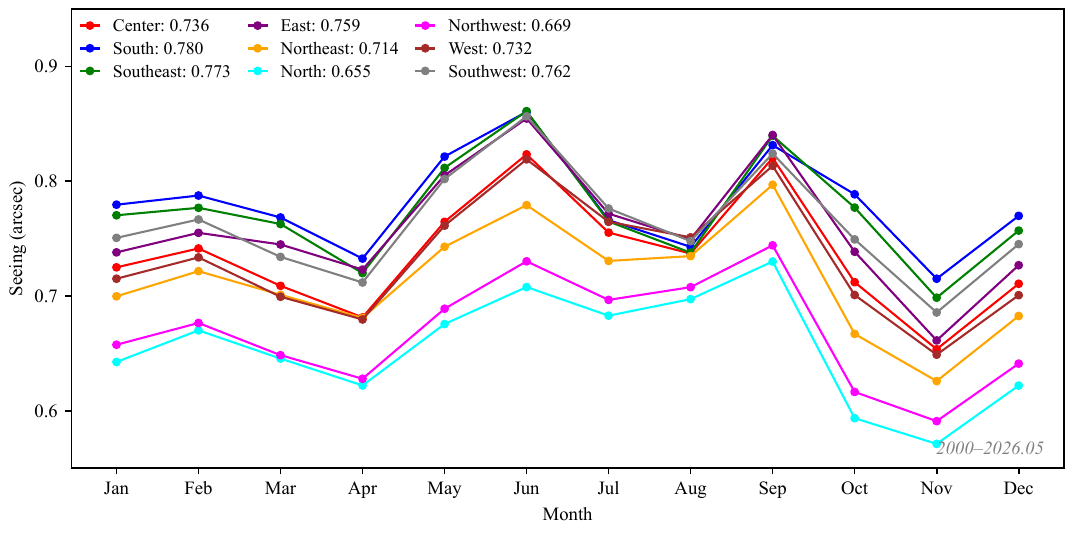}
 \caption{Monthly mean nighttime seeing at $\lambda=500$~nm for the nine sites from 2000 to 2026.}
\label{fig:general_12}
\end{figure*}

The calculated results are shown in Figure~\ref{fig:general_12}, where the horizontal axis represents the month and the vertical axis represents seeing in units of arcseconds (arcsec). As can be seen from the figure, the seeing in this region exhibits a pronounced north--south difference: the three southern sites constitute the group with the poorest seeing values, while the northern sites have the best observing conditions. The central Platform C has an annual mean seeing of 0.736 arcsec, which is at an intermediate level across the entire region.

The seasonal variation trends are consistent across all sites. seeing rises to its highest level of the year in June and September, when upper-level atmospheric disturbances are most intense, and falls back to its lowest annual range from October to November, which is the period with the weakest turbulence throughout the year.

Overall, the seeing integrated only above the site level is not affected by differences in surface elevation, yet it still exhibits a stable north-superior/south-inferior spatial differentiation pattern, corroborating the patterns identified in the previous sections.

 \subsection{Validation against ground-based measurements}\label{comparisons}

When applied the goal of current work, apart from the advantages of data coverage in time and geographic scales the MERRA-2 data has obvious limitations due to limited vertical resolution. The turbulence profile in the light path of star lights normally features fine, strong layers~\citep{chun2014extremely, els2009thirty, tokovinin2005optical} that cannot be resolved by the 72 model levels of the MERRA-2 reanalysis~\citep{gelaro2017merra2}. Thus the result derived from the reanalysis data should be crossly checked with other regular on site measurements. As direct assessments of turbulence profile are not available, we can use MASS and DIMM monitoring data sets as well as radiosonde balloon data to indirectly check the integrated quantities; however, considering that radiosonde balloons are relatively scarce, with only one day overlapping in time with the other data sets, only the MASS and DIMM data are selected for the cross-validation, and the single day of radiosonde balloon data is used later only as an additional illustration of the profile shape rather than for quantitative validation.

\begin{figure*}
\centering
\includegraphics[width=\textwidth]{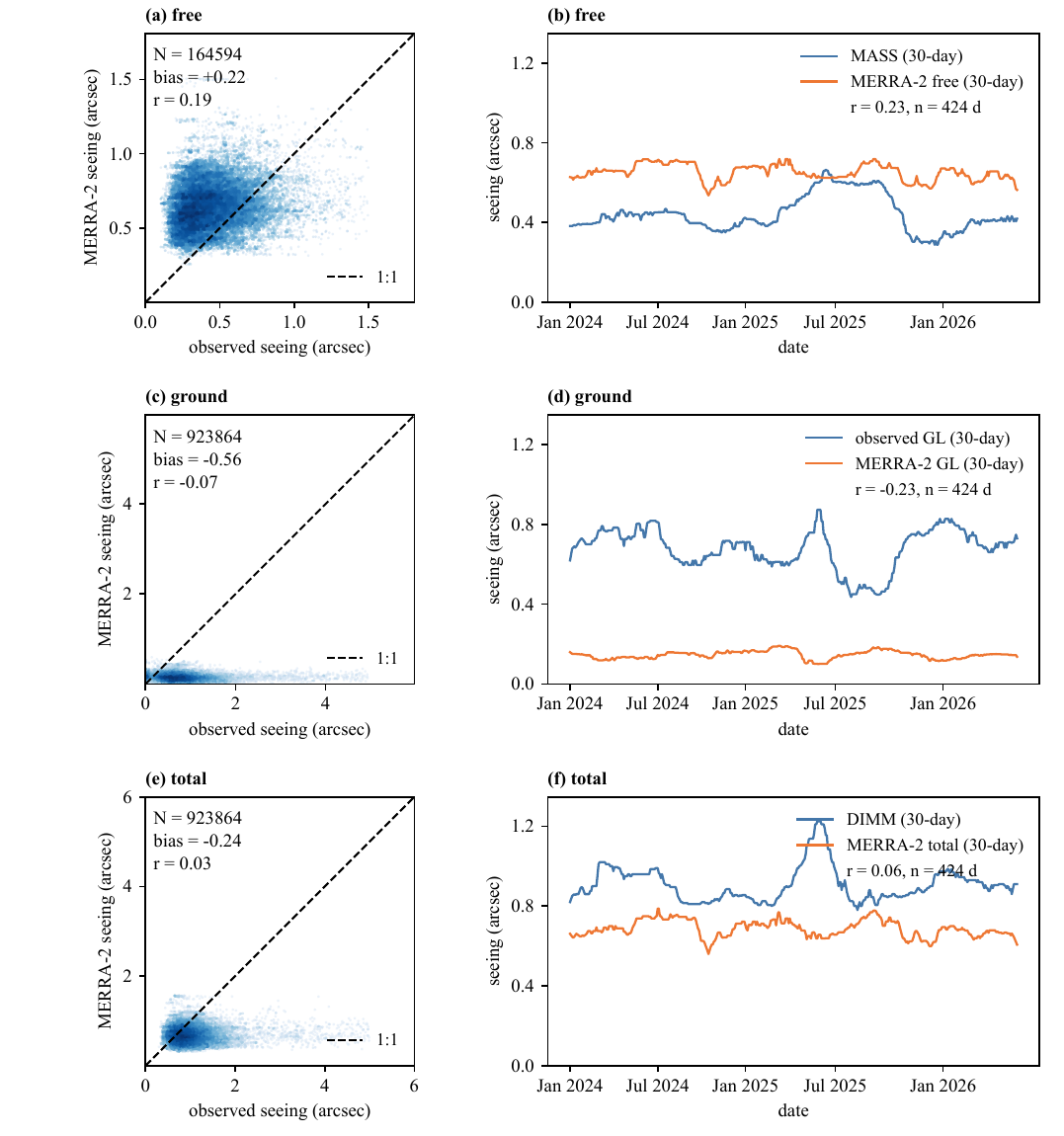}
\caption{Validation of the MERRA-2 derived seeing against simultaneous MASS and DIMM measurements at Platform C over the 2024--2026 overlap period (28 months).}
\label{fig:validation}
\end{figure*}

We have two instruments operating at Platform C: a MASS instrument~\citep{ni2009mass}, which resolves the optical turbulence into six layers at $0.5,1,2,4,8$ and $16$~km above the instrument and thus provides the free-atmosphere seeing above $\sim0.5$~km, and a DIMM, which measures the total integrated seeing including the ground layer. The instruments were started at the early stage of site construction at different times, the operation and data taking of both were not very stable because of site activities going on all the time. The duty cycle of the site quality monitoring including the two instruments is affected by power down or blast of mountain top within the observatory. We compare the reanalysis against \emph{every} on-site measurement: for each MASS and DIMM epoch the 3-hourly MERRA-2 $C_n^2$ profiles are linearly interpolated in time to the epoch of the observation, and the interpolated MERRA-2 $C_n^2$ profile is integrated above $562.5$~hPa (approximately corresponding to the lowest altitude that MASS can measure) to obtain the free-atmosphere seeing and above $600$~hPa to obtain the total seeing. Since the turbulence integral is additive along the line of sight, the total and free-atmosphere seeing combine in the $5/3$ power, and the ground-layer (GL) seeing is obtained per epoch as
\begin{equation}
\varepsilon_{\rm GL}=\left(\varepsilon_{\rm tot}^{5/3}-\varepsilon_{\rm FA}^{5/3}\right)^{3/5},
\label{eq:glseeing}
\end{equation}
where all three quantities refer to the same wavelength ($\lambda=500$~nm) and the same air mass~\citep{tokovinin2007accurate,kornilov2007combined}; epochs for which the difference is negative are set to zero. The point-by-point comparison is shown in the left column of Fig.~\ref{fig:validation}, and the variation of the respective parameters with date in the right column.

\begin{figure*}
\centering
\includegraphics[width=\textwidth]{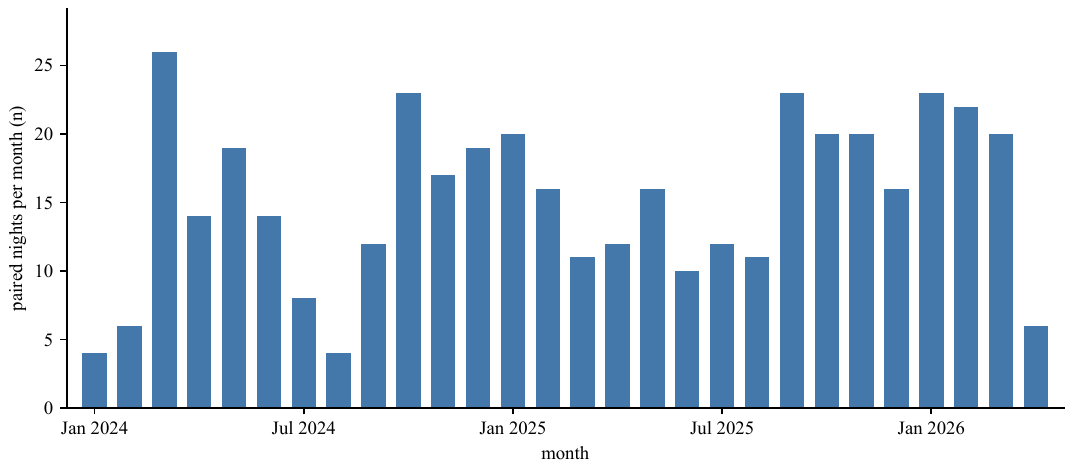}
\caption{Number of paired nights (DIMM and MASS both available) per calendar month in the overlap period (2024 January--2026 April).}
\label{fig:nightscount}
\end{figure*}

In the free atmosphere (Fig.~\ref{fig:validation}a), MERRA-2 overestimates the measured seeing at every epoch; because consecutive epochs within a night are not independent, the correlation statistics are computed on the per-night medians of the $n=424$ paired nights (Fig.~\ref{fig:validation}b): the night-to-night correlation is weak ($r=0.23$; 95\% bootstrap interval $0.14$--$0.33$), i.e.\ the reanalysis reproduces the median free-atmosphere state but not its variability. The relative bias of the total seeing (Fig.~\ref{fig:validation}e) has the opposite sign, $-27\%$, and the night-to-night correlation of the same set of per-night medians (Fig.~\ref{fig:validation}f) essentially vanishes ($r=0.06$; 95\% interval $-0.04$--$0.16$). The point-by-point correlation over all epochs is likewise very low (for the free atmosphere, the ground layer and the total seeing, $r=0.19$, $-0.07$ and $0.03$, respectively; Fig.~\ref{fig:validation}a,c,e). This shows that MERRA-2 has essentially no skill in predicting the night-to-night total seeing, capturing only about $22\%$ of the ground layer that dominates it (Table~\ref{keystats}; Fig.~\ref{fig:validation}c,d). This follows directly from the smoothed model topography: the centre grid point lies $\sim1300$~m below the Platform C summit (Tables~\ref{tab:grid_points} and~\ref{keystats}), so the model has essentially no resolved boundary layer above the site.

In Fig.~\ref{fig:validation} both the DIMM and the MASS data show a fluctuation peak around May 2025, whereas MERRA-2 does not exhibit this feature; this is mainly because the observing base was under construction at that time, which made the measured seeing biased high.

\begin{figure*}
\centering
\includegraphics[width=\textwidth]{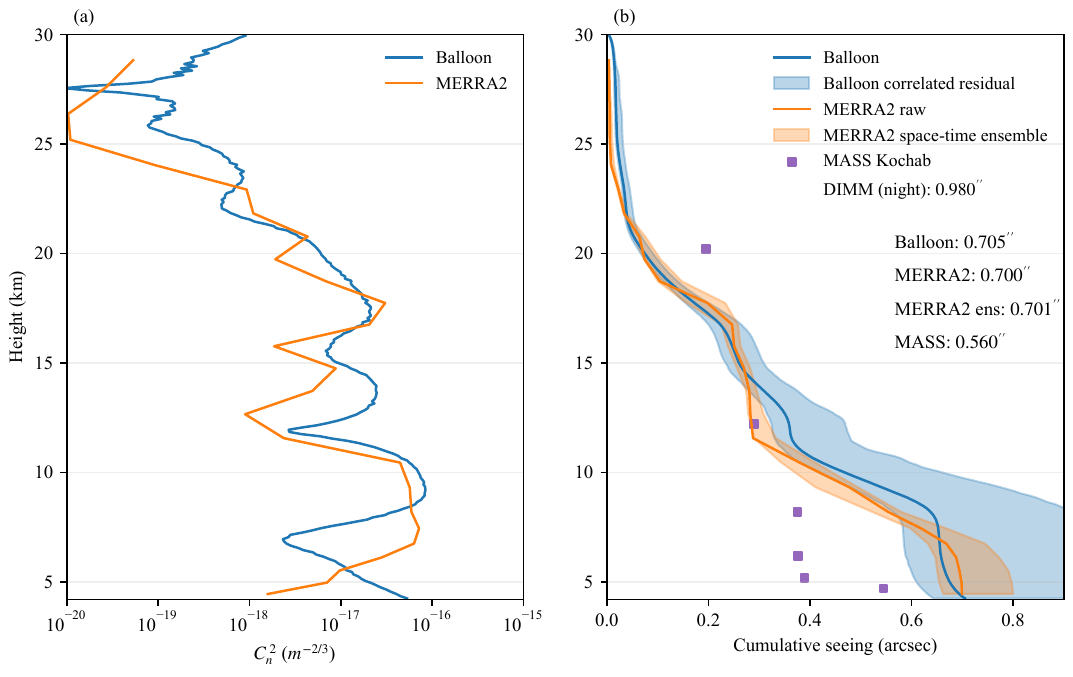}
\caption{Comparison of $C_n^2$ (panel a) and of the cumulative seeing (panel b) derived from the MERRA-2 reanalysis dataset and from a nearby radiosonde balloon sounding at the Lenghu site on 28 June 2025 (balloon release at 23:23~UTC, i.e.\ 07:23 Beijing time on the following morning).}
\label{fig:general_16}
\end{figure*}

The only radiosonde balloon profile that overlaps in time with MERRA-2 is compared with the MERRA-2 profile and placed in the context of the simultaneous MASS measurements of the same night, as shown in Fig.~\ref{fig:general_16}.

This figure presents the vertical structures of the refractive index structure constant C$_n^2$ of atmospheric turbulence and seeing at Platform C of the Lenghu Astronomical Observation Base on 28 June 2025. The radiosonde balloon was released at 23:23 UTC at Platform C. For the MERRA-2 reanalysis the time slice closest to the balloon release, i.e.\ 00:00 UTC, is selected, and data from the nearest grid point (93.75$^\circ$E, 38.5$^\circ$N) to Platform C are extracted for calculation. The left panel shows the vertical profile of the atmospheric refractive index structure constant C$_n^2$. The solid blue line is the sounding profile, computed with the same method as the reanalysis; it can be seen that the balloon profile resolves a series of thin, strong layers that the smoothed MERRA-2 profile cannot reproduce, illustrating the limitation of the reanalysis discussed in Section~\ref{comparisons}. The right panel displays the cumulative seeing. The two shaded bands in the figure answer two different questions: the orange band is obtained by recomputing the profiles for the nine neighbouring grid points and three consecutive 3-hourly time slices (27 realisations in total), and the 5th--95th percentiles of these 27 realisations form the orange band, which measures the error introduced by the choice of the neighbouring grid points; the raw resolution of the sounding is about 6~m, and it must be smoothed to about 1~km, comparable to that of the model, before it can be compared with the reanalysis on the same footing. The blue band is then obtained by adding a vertically correlated random field to the sounding's own $\log C_n^2$ residuals (correlation length $L_c\approx0.5$~km, amplitude $\sigma(z)$ measured from those residuals), making 1000 realisations and taking the 5th--95th percentiles; it quantifies the sensitivity of the column integral to structure below the effective vertical resolution of the reanalysis ($\sim$1~km, the spacing of the MERRA-2 layers), i.e.\ it reflects the sensitivity of the column integral to structure below about 1~km. Neither band contains the $C_n^2$ parameterisation constants: these constants are identical in all realisations and cancel when differences are taken, so that the bands describe only the response of the column integral to the shape of the profile. The purple squares are the layered cumulative seeing measured by MASS on the star Kochab during UTC 15:32--21:19, covering the free atmosphere above $\sim$0.5~km; because radiosondes are released at fixed (`synoptic') times, this sounding (23:23~UTC, following the end of the MASS window by about two hours) is the closest available profile to the MASS and DIMM observations of that night. The inset of panel (b) additionally lists the full-column total seeing from DIMM ($0.980''$), which is a full-column quantity and has no height dependence, and therefore does not appear as a point on the curves; combining it with the MASS value in the $5/3$ power gives a ground-layer contribution of about $0.73''$ for that night, consistent with the statistical result of Section~\ref{comparisons}. With the same technique, we have also calculated a few critical parameters that are normally required for site qualification from the same measurements. The key parameters calculated to facilitate a comparison with other sites are presented in Table~\ref{keystats}.

It should be noted that the unmarked turbulence parameters for Cerro Armazones and Mauna Kea 13N listed in Table~\ref{keystats} are directly extracted from Table 2 of~\citet{schock2009tmt}; here the MERRA-2 windows are aligned with the instrument operation periods documented in the TMT site-testing database, and Lenghu C is a newly analysed site added in the present work.

All statistical results for Lenghu C are derived from two data sources: on-site DIMM/MASS measurements and MERRA-2 atmospheric reanalysis data. The DIMM and MASS seeing values are the medians of the per-night medians over 2024 January 17 to 2026 April 6, restricted to the $n=424$ nights on which both instruments were available (Fig.~\ref{fig:nightscount}); the free-atmosphere, total and ground-layer values thus refer to the same nights and are the same statistics used in the validation of Section~\ref{comparisons}. The coherence time ($\tau_0$) marked with a dagger ($^\dagger$) is derived by combining the full MASS turbulence profile with temporally matched MERRA-2 wind speed profiles; the MERRA-2 values for Lenghu C are computed over the same 424 paired nights, evaluating at a wavelength of $500$~nm the 18:00 UTC (approximately 02:00 local Beijing time) reanalysis profiles, which fall close to local midnight, as the night representative, as in Section~\ref{comparisons}. The MERRA-2 values for Cerro Armazones and Mauna Kea 13N are likewise computed within the TMT project operation periods of~\citet{schock2009tmt} (2004 November--2009 September for Armazones and 2005 June--2008 May for Mauna Kea 13N, as documented in the TMT site-testing database). The values are taken from the reanalysis time slices closest to local midnight at each site (06:00 UTC at Cerro Armazones and 09:00 UTC at Mauna Kea 13N; 18:00 UTC, i.e.\ 02:00 local time, at Lenghu C); the $\theta_0$ values are integrated above the lower boundary of the MASS measurements (562.5~hPa at Lenghu C, and 653~hPa and 575~hPa at Cerro Armazones and Mauna Kea 13N, corresponding to the same $\approx$0.5~km offset above each site), while the $\tau_0$ values at Cerro Armazones and Mauna Kea 13N are integrated above the site elevation (695~hPa and 612~hPa) because their literature values combine the free atmosphere with the ground layer, and the Lenghu C value is integrated above 562.5~hPa.

\begin{table*}
\centering
\begin{threeparttable}
\caption{Key profile parameter statistics of different sites.}\label{keystats}
\begin{tabular}{llccc}
\hline
Parameter & Data source & Lenghu C & Armazones & Mauna Kea 13N\\
\hline
Elevation (m) & & 4200 & 3064 & 4050 \\
Total seeing ($''$) & DIMM & 0.89 & 0.64 & 0.75 \\
Free atmosphere seeing ($''$) & MASS & 0.44 & 0.43 & 0.33 \\
GL seeing 7--500~m ($''$) & D-M & 0.68 & 0.35 & 0.54 \\
Isoplanatic angle, $\theta_0$ ($''$) & MASS & 2.05/1.72$^*$ & 2.04/1.32$^*$ & 2.69/1.52$^*$ \\
Coherence time, $\tau_0$ (ms) & M+D & 4.32$^\dagger$/1.92$^*$ & 4.6/2.38$^*$ & 5.1/4.09$^*$ \\
\hline
\end{tabular}
\begin{tablenotes}
\item[*] Values marked with an asterisk are calculated in this work from the MERRA-2 reanalysis profiles, evaluated at $\lambda=500$~nm.
\item[$\dagger$] Coherence time for Lenghu C derived by combining the MASS turbulence profile with the MERRA-2 wind profile.
\end{tablenotes}

\end{threeparttable}

\end{table*}

As shown by the statistical results in Table~\ref{keystats}, there is a clear systematic discrepancy between the direct site measurements and the values derived from MERRA-2 reanalysis, manifesting itself as an overestimated free-atmosphere seeing and underestimated $\theta_0$, $\tau_0$ and total seeing. The reasons for this are mainly the following. First, topographic smoothing and spatial resolution: because MERRA-2 is a global reanalysis model with a relatively coarse horizontal resolution, and astronomical observatories such as Lenghu, Armazones and Mauna Kea are invariably situated on isolated mountain peaks, the model topography significantly ``flattens'' these sharp peaks, smoothing out the local thermal inversions and stable boundary layers. Second, parameters such as the isoplanatic angle ($\theta_0$) depend heavily on high altitudes ($z^{5/3}$ weighting), and these parameterisation schemes tend to overestimate the thickness and intensity of turbulent layers in high-shear regions, thus systematically underestimating $\theta_0$. Moreover, MERRA-2 provides instantaneous three-hourly snapshots, unlike MASS, which directly measures high-frequency stellar scintillation to retrieve the optical turbulence profile, and therefore cannot resolve the transient, highly stable, ``laminarised'' flows that occur at the site under optimal observing conditions, so that the calculations show a higher turbulent ``background'' state.

 \section{Conclusion}
\label{sec:style_5}

Taking the grid point closest to the Lenghu Astronomical Observation Base as the centre, a total of nine sites were selected---one central site and eight surrounding sites located 155--167~ km away in the east, south, west, north, northeast, southeast, northwest, and southwest directions. Based on the MERRA-2 dataset covering the period from 2000 to 2026, a systematic analysis was conducted on the spatial distribution and seasonal variation characteristics of the temperature field, wind field, atmospheric refractive index structure constant, atmospheric coherence parameters, and optical seeing.

\begin{enumerate}
\item Seasonal Influence of Temperature and Wind Speed.

The temperature and wind fields in the Lenghu region show clear seasonal variations, particularly the differences between summer and winter. Winter temperatures are lower, and wind speeds are higher, while summer temperatures are higher, and wind speeds are lower. The spring and autumn seasons exhibit smaller temperature and wind speed fluctuations, resembling transitional seasons.
\item Seasonal Variation of Turbulence Intensity.

In the Lenghu region, the intensity of atmospheric turbulence shows distinct seasonal variation. In summer, the turbulence intensity at low altitudes is higher, especially in the middle and lower troposphere (600--400 hPa), due to strong surface heating and intense convective activity. In winter, the turbulence intensity is lower, particularly at high altitudes, where atmospheric turbulence activity is significantly reduced. The turbulence intensity in spring and autumn is intermediate, exhibiting transitional characteristics between summer and winter.

\item Influence of Meteorological Factors on Turbulence.

The temperature and wind fields are closely related to turbulence intensity. The intensity of low-altitude turbulence is usually associated with surface heating and atmospheric convection activity, while high-altitude turbulence intensity is influenced by pressure and temperature gradients. The stable stratification of the lower stratosphere suppresses turbulence activity, so that between 100 and 200 hPa the turbulence intensity is significantly reduced.

\item Turbulence Assessment for Site Selection.

The atmospheric turbulence characteristics in the Lenghu region provide valuable reference for the site selection of large-aperture telescopes. By analysing turbulence intensity in different directions, significant differences were found between the north and south. The northern area is relatively stable with lower turbulence intensity, making it suitable for high-precision observations. In contrast, the southern area has stronger low-level turbulence, which is less favourable for high-resolution imaging but remains acceptable for wide-field, seeing-limited observations. The turbulence differences in the east and west directions are relatively small.

\item Application of Site Evaluation Results.

 The results provide a theoretical basis for the construction of the Lenghu observatory and for telescope placement and science planning; absolute design parameters, however, still require dedicated on-site measurement. The turbulence profiles, meteorological parameters, and turbulence characteristics obtained from this study offer important references for selecting telescope placement, planning scientific tasks, and the long-term operation of the site.

\item Validation and Applicability of the Reanalysis.

A systematic comparison of the reanalysis data over the 28-month overlap period (2024 January--2026 April) against every simultaneous MASS and DIMM measurements at Platform C shows that the MERRA-2 captures only a small portion of the ground-layer turbulence, and overestimates free-atmosphere seeing median, with weak night-to-night correlation, while underestimating the total seeing with negligible night-to-night correlation, giving little of the observed seasonal amplitude of the free atmosphere (as discussed in \ref{comparisons}). The reanalysis data possesses obvious limitations in solving details of profiles, but it is still suited to relative comparisons between locations and to multi-year climatological characterisation; For any quantitative analysis the turbulence profiles over a site, direct on-site measurements are needed.

\end{enumerate}

\section*{Acknowledgments}

This work is supported by the National Natural Science Foundation of China (grants 12233009, 42222408 and 12373035), and the Ministry of Science and Technology (MOST grant 2023FY101100).

\section*{Data Availability}

The MERRA-2 reanalysis data used in this study are publicly available from the NASA Goddard Earth Sciences Data and Information Services Center (GES DISC) at \url{https://disc.gsfc.nasa.gov/}, in the collection of instantaneous three-dimensional assimilated meteorological fields on the model levels (\texttt{inst3\_3d\_asm\_Nv}). The MASS and DIMM monitoring data from the Lenghu site are publicly available at \url{http://lenghu.china-vo.org}. Topographic elevations are taken from the CGIAR-CSI SRTM 90~m Digital Elevation Database v4.1, derived from the USGS/NASA SRTM data and distributed at \url{https://srtm.csi.cgiar.org}. The derived data products underlying this article (the monthly mean profiles, the per-night seeing series and the statistical tables) are available from the corresponding author upon reasonable request.

\bibliographystyle{mnras}
\bibliography{reference}

\bsp	
\label{lastpage}
\end{document}